\documentstyle[aps,prl,multicol]{revtex}
\begin{document}
\title{Growth enhanced surface diffusion and elastic instability on
amorphous solids}
\author{Martin Rost}
\address{Helsinki Institute of Physics, P.O.\ Box 9, 00014 University
of Helsinki, Finland \\
and Laboratory of Physics, P.O.\ Box 1100, 02015 HUT, Espoo, Finland}
\date{\today}
\maketitle
\begin{abstract}
A continuum model for growth of solids is developed, considering
adatom deposition, surface diffusion, and configuration dependent
incorporation rate. For amorphous solids it is related to surface
energy densities. The high adatom density leads to growth enhanced
dynamics of (a) Mullins' classical equation [J.\ Appl.\ Phys.\ {\bf
28}, 333 (1957)] without, and (b) of the
Asaro-Tiller-Grinfeld-Srolovitz instability with lateral stress in the
growing film. The latter mechanism is attributed to morphologies found
in recent experiments.
 
\end{abstract}
\pacs{05.20.-y, 81.15.Aa, 68.55.-a, 62.20.Dc, 61.43.Dq}

\begin{multicols}{2}
The theoretical approach to kinetic roughening \cite{Krug91}, due to
its roots in Statistical Mechanics, relates various growth processes
on solids to so called universality classes. Their distinction allows
to identify essential properties for the large scale morphology beyond
the microscopic mechanisms, say certain symmetries or conservation
laws, which manifest themselves in terms of universal scaling
exponents.

One such important property is mass conservation suppressing adatom
loss to the surrounding space, which allows for surface relaxation
only through transport along the surface. The total solid mass then is
always equal to its initial value plus the integrated incoming flux.

Molecular Beam Epitaxy (MBE) garantees such conditions, and growth
parameters like for instance flux intensity, temperature, chemical
composition of the adsorbate can be easily and precisely controlled
\cite{BuchMBE}. It has been widely used and refined for growth of
semiconductor and metal crystals. However, under typical conditions
the diffusion length and average crystal step distance are quite
large, imposing their own features to the morphology
\cite{Villain91,Krug97}, such that any theoretically predicted
asymptotic scaling regime lies beyond the reach of experimental
observation.

On the other hand it may become observable in growing amorphous
substances where it seems plausible that intrinsic lengthscales remain
small. Some experiments have been performed in recent years looking
for kinetic roughening on amorphous solids
\cite{MoskeHabil,Moske97,Moske99,Geyer97,Yang96,Lutt97}. It turns out
that their theoretical interpretation in terms of standard continuum
\cite{Mullins59} and discrete \cite{WolfVillain,DasSarma} models
leaves two main open questions which are addressed in this
letter. First, far from equilibrium the kinetics on the surface are
different from relaxation by thermally activated adatoms leading to
Mullins' continuum equation \cite{Mullins59}. Second, in some
experiments there appears a very pronounced intermediate lateral
lengthscale together with a mounding instability
\cite{MoskeHabil,Moske97,Moske99,Geyer97} whose origin was
unclear. Obviously it cannot be due to step edge barriers as is the
case on crystal surfaces \cite{Villain91,Ehrlich66}. Another suggested
mechanism, deflexion of incoming particle trajectories or ``steering''
\cite{Moske99}, is in some cases relevant at grazing incidence but
negligible for normal beams \cite{Dijken}.  

Here a new continuum approach to growth with surface diffusion is
developed resulting in a flux dependent coefficient for
surface evolution (see Eqs.\ (\ref{lin_surf}) and (\ref{final_lin})),
the counterpart to equilibrium relaxation Eq.\ (\ref{mullins_eq})
\cite{Mullins59}. The mounding instability is shown to be due to
lateral elastic stress in the growing solid
\cite{MoskeHabil,Geyer97,Moske89,Dina92}, which is responsible for the
well known Asaro-Tiller-Grinfeld-Srolovitz (ATGS) instability
\cite{Asaro,Grinfeld,Srolovitz,Politi99,Spencer91}.
Surface modulations allow for relaxation of a laterally stressed
solid. For large enough wavelengths this energy gain overcomes the
cost in additional capillary energy, disfavouring and destabilizing
a flat surface. Due to the nonequilibrium kinetics it appears in a new
light: Its driving force is an equilibrium, the dynamical evolution a
nonequilibrium phenomenon. In the following calculations the surface
energetics account well for the observed wavelength (the size of
emerging mounds), and the nonequilibrium kinetics yield the linear
growth rate of the instabilty.

To begin, the continuum model of equilibrium relaxation is briefly
recalled \cite{Mullins59}. Surface configurations are described by a
space and time dependent height field $H({\bf x},t)$, neglecting
overhangs and voids. The driving force for equilibrium relaxation is
the surface free energy, ${\cal F} = \int \! d^2x \, \gamma \sqrt{1 +
(\nabla H)^2}$. If an atom is added somewhere at the surface this
total area may change, resulting in a extra energetic contribution
$\mu = - \Omega \gamma \nabla^2 H$ \cite{Linear}. $\mu$ is often
called chemical potential and $\Omega = a^3$ denotes the atomic
volume. Spatial variations in $\mu$ bias the adatom diffusion and
create a macroscopic mass exchange. Local mass (and volume)
conservation impose a continuity equation for the surface dynamics,
which for $|\mu| \ll k_B T$ \cite{Linear} reads 
\begin{equation}
\label{mullins_eq}
\partial_t H = - a \nabla \cdot D \frac{\Omega \gamma}{k_B T} \nabla
\nabla^2 H \equiv - \kappa (\nabla^2)^2 H.
\end{equation}
A reasonable assumption for the adatom diffusion coefficient is $D = 
a^2 \! / \tau_0 \, e^{-\varepsilon_{\rm eff}/k_B T}$, an Arrhenius
form with some activation barrier $\varepsilon_{\rm eff}$ (which on a
random surface comes from an average over waiting times at each site
\cite{Kehr}) and an ``attempt frequency'' $1/\tau_0$
\cite{SmilauerVvedensky}. Eq.\ (\ref{mullins_eq})
predicts an exponential decay with rate
$-\kappa k^4$ for fluctuations with wavenumber $k$, which has been
veryfied experimentally \cite{Bonzel}. On a phenomenological basis
Eq.\ (\ref{mullins_eq}) has also been applied to growth with
relaxation by curvature driven surface diffusion, where under
deposition (``shot'') noise it results in a power spectrum of height
fluctuations $S({\bf k},t) \equiv \langle \vert H({\bf k},t) \vert^2
\rangle \propto \vert {\bf k} \vert^{-4}$ \cite{Krug97}. Of course
under growth the coefficient $\kappa$ takes a different form as above
in Eq.\ (\ref{mullins_eq}) which is adressed next.

On a growing solid there is a constant supply of mobile adatoms out of
the beam hitting the surface. The number of thermally activated
adatoms, responsible for equilibrium relaxation, now becomes
negligible. Here an ansatz is presented where the incorporation rate
$I \{ H \} $ depends on the surface configuration $H({\bf x},t)$ and
connects it thus to the adatom density $\rho({\bf x},t)$,
\begin{eqnarray}
\label{ansatz}
\partial_t H & = & I \{ H \} \Omega \rho \\
\partial_t \rho & = & \nabla \cdot D\{ H,\rho \} \nabla \rho - I \{ H
\} \rho + F/\Omega \nonumber.
\end{eqnarray}
Note that $H+\Omega \rho$ is a conserved quantity, increasing by the
average growth velocity $F$. The problem is to find
the dependence $I \{ H \} $, but before turning to that point its
r\^ole in a phenomenological equation as (\ref{mullins_eq}) shall be
worked out. 

Eqs.\ (\ref{ansatz}) are expanded around the ``flat'' growing surface
$H_0(t) \equiv H_0 + Ft$, on which the incorporation rate takes some
value $I_0$, and around the average adatom density $\rho_0 \equiv
F/(\Omega I_0)$. Small deviations $h({\bf x},t) \equiv H({\bf x},t) -
H_0(t)$ and $c({\bf x},t) \equiv \Omega(\rho({\bf x},t) - \rho_0)$
obey to linear order
\begin{eqnarray}
\label{lin_ansatz}
\partial_t h & = & - \frac{F}{I_0} I_1 \ast h + I_0 c \\
\partial_t c & = & \frac{F}{I_0} I_1 \ast h + (D \nabla^2 - I_0) c,
\nonumber
\end{eqnarray}
where $D \! \equiv \! D\{ H_0(t),\rho_0 \}$. $I \{ H \} \! - \! I_0 \!
\approx \! I_1 \ast h$ is approximated linearly (not necessarily local
in space) and will be evaluated in its Fourier transform $I_1 ({\bf
k})$.

In the long wavelength limit, $D k^2 \ll I_0$ (beyond the diffusion
length $\ell_d$, the typical distance an adatom moves before
incorporation), and also $\vert I_1({\bf k}) \vert F/I_0 \ll I_0$
\cite{IncoComm}, the eigenvalues of (\ref{lin_ansatz}) are
\begin{eqnarray}
\label{eigenv}
\lambda_1 & = & - \frac{DF}{I_0^2} k^2 I_1({\bf k}) \nonumber \\
\lambda_2 & = & - I_0
\end{eqnarray}
with relative errors $O(Dk^2/I_0) + O(F \vert I_k({\bf k}) \vert
/I_0^2)$. $\lambda_2$, remaining finite in the long wavelength limit,
rules the enhanced (diminished) incorporation under higher (lower)
adatom density. The eigenvector of $\lambda_1$ mainly lies in
direction of $h$, so in the considered large scale limit the
linearization of Eqs.\ (\ref{ansatz}) turns into
\begin{equation}
\label{lin_surf}
\partial_t h({\bf k},t) = - \frac{DF}{I_0^2} k^2 I_1({\bf k}) h({\bf
k},t),
\end{equation}
a non-equilibrium version of Eq.\ (\ref{mullins_eq}).

On crystal surfaces $I_1({\bf k})$ depends essentially on the step
configurations, but on amorphous solids it is proportional to
variations in energy density at the surface, and can be derived in a
mean field type approach. For a diffusing adatom the amorphous surface
consists of potential wells of various depths. They have an average
distance $a$ from each other and a probability distribution of depths
$n(\varepsilon/\varepsilon_0)/\varepsilon_0$, where the energy scale
$\varepsilon_0$ is explicitely included in the notation. In a simple
view of the nonequilibrium growth process an atom gets deposited,
diffuses, and is finally incorporated into the solid in a sufficiently
deep potential well. Dimer formation or other types of nucleation can
be neglected if sticky sites are denser than adatoms --- a condition
to be verified a posteriori. An adatom ``sticks'' to a site, if it
cannot escape until it is buried by the further growing solid. Thus
the depths $\varepsilon$ of ``sticky'' potential wells fulfill $\tau_0 
e^{\varepsilon/k_B T} > a/F$. Assuming an exponential distribution of
energy depths, $n(\varepsilon/\varepsilon_0) =
\exp(-\varepsilon/\varepsilon_0)$ \cite{CommExp}, this yields a
relative proportion of sticky sites
\begin{equation}
\label{sticky_dens}
\frac{a^2}{\ell_d^2} = \left( \frac{F \tau_0}{a} \right)^{k_B
T/\varepsilon_0},
\end{equation}
which also defines the diffusion length $\ell_d$, because the fractal
dimension of a random walk is two and the adatom can ``fully explore''
its surroundings on the surface. The average incorporation rate is the
inverse of the time needed to hit a sticky site, so $I_0 =
D/\ell_d^2$.

As seen in the context of Eq.\ (\ref{mullins_eq}) deviations from a
flat surface change the energy density and the chemical potential near
the surface. A natural way to account for this is shifting the energy
scale $\varepsilon_0 \to \varepsilon_0 - \mu = \varepsilon_0 - \Omega
{\cal E}$. A positive additional energy density ${\cal E}$ lowers the
depth of potential wells encountered by the adatoms, and changes the
incorporation rate by
\begin{equation}
\label{inc_change}
I - I_0 =  - I_0 \log \frac{a}{F \tau_0} \; \frac{k_B T}{\varepsilon_0}
\frac{\Omega {\cal E}}{\varepsilon_0} = - 2 I_0 \log
\frac{\ell_d}{a} \; \frac{\Omega {\cal E}}{\varepsilon_0}. 
\end{equation}
The change is proportional to the relative energy change for
adatoms. Flux and temperature enter via $\ell_d$ and $I_0$.

Now the surface energetics have to be evaluated in order to obtain
$I_1({\bf k})$ in Eq.\ (\ref{lin_surf}) via (\ref{inc_change}). As
above in Eq.\ (\ref{mullins_eq}) also here the change in surface free
energy enters, so there must be a capillary contribution \cite{Linear}
\begin{equation}
\label{e_cap}
{\cal E}_{\rm cap}^{\rm lin} ({\bf x},t) = - \gamma \nabla^2 h({\bf
x},t).
\end{equation}

The other important part of ${\cal E}$ comes from elastic deformations
of the growing solid. In the experiments considered here amorphous
metallic glasses (i.e.\ the alloy Zr$_{\rm 65}$Al$_{\rm 7.5}$Cu$_{\rm
27.5}$) grow under lateral expansive stress the origin of which is
not fully clarified \cite{MoskeHabil,Geyer97,Moske89,Dina92}. It
builds up during growth and reaches a constant level at a film
thickness of about 50 nm. The order of magnitude is then 1 GPa,
corresponding to a lateral deformation of $\alpha \approx 0.1$ to $1$
\%, given Young's modulus to be roughly $E \approx 10^2$ GPa
\cite{MoskeHabil,Moske89,Dina92}. Applying the linear relations of
stress and strain for small deformations \cite{LandauLifshitz} one
can calculate the strain and stress tensors in the film with a flat
surface, $u_{ij}^0$ and $\sigma_{ij}^0$, as well as the elastic energy
density ${\cal E}_{\rm el}^0 = \sigma_{ij}^0 u_{ij}^0/2$.

Changes in shape are slow compared to mechanical balancing
inside the body,
so the strain and stress fields follow the surface 
configuration quasistatically. Surface variations can be seen as
perturbing $\sigma_{ij}^0$ by an additional stress field $\tau_{ij}$
with boundary conditions $\tau_{iz} \! = \! - \partial_i h E \alpha /
(1 \! - \! \sigma)$ for $i \! = \! x,y$ and $\tau_{zz} \! = \! 0$ to
linear order in $\nabla h$ ($\sigma$ without indices denotes the
Poisson number). Green's function corresponding to the geometry of the
body \cite{LandauLifshitz} yields the perturbative $\tau_{ij}$ and
corresponding strain $w_{ij}$ throughout the solid. In view of Eq.\
(\ref{inc_change}) only the corrections to the elastic energy density
${\cal E}_{\rm el}^0$ at the surface to linear order in
$\tau_{ij}$ and $w_{ij}$ (and therefore $\nabla h$) are needed,
\begin{equation}
\label{E_el_misfit}
{\cal E}_{\rm el}^{\rm lin} ({\bf x},t) = \frac{\alpha^2}{\pi} \frac{E}{1-2
\sigma} \int d^2 x' \;
\frac{ ({\bf x}-{\bf x'})\cdot \nabla h({\bf
x'},t)}{\vert{\bf x}-{\bf x'}\vert^3}.
\end{equation}

Curved parts of the surface also compress or elongate the solid,
contributing to the energy density by
\begin{equation}
\label{E_el_surftens}
{\cal E}_{\rm el'}^{\rm lin}({\bf x},t) = - \sigma
\frac{1-2\sigma}{1-\sigma} \alpha \gamma \; \nabla^2 h({\bf x},t).
\end{equation}
The derivation of these elastic energy densities will be presented
elsewhere in more detail \cite{Elasti}. Here the next step is to
gather Eqs.\ (\ref{e_cap}), (\ref{E_el_misfit}), and
(\ref{E_el_surftens}) to the Fourier transform of the energy density
variations at the surface,
\begin{eqnarray}
\label{E_lin_total}
{\cal E}^{\rm lin}({\bf k},t) & = & B({\bf k}) \; h({\bf k},t)  \\
= \Biggl[ \! \Biggl( 1 & + & \sigma \alpha \frac{1 \! - \! 2\sigma}{1
\! - \! \sigma} \Biggr)  \gamma  k^2 - \frac{E}{\pi} \; \alpha^2
\frac{1 \! + \! \sigma}{1 \! - \! \sigma} \vert {\bf k} \vert \Biggr]
h({\bf k},t), \nonumber
\end{eqnarray}
which is local in ${\bf k}$. With $B({\bf k})$ defined above, the
linear surface evolution in Eq.\ (\ref{lin_surf}) including capillary
and elastic effects becomes
\begin{equation}
\label{final_lin}
\partial_t h({\bf k},t) = - 2 \ell_d^2 \log \frac{\ell_d}{a} \; F
\frac{\Omega B({\bf k})}{\varepsilon_0} k^2 h({\bf k},t).
\end{equation}
In particular, without any elastic effects the kinetic coefficient in
Eq.\ (\ref{mullins_eq}) is $\kappa = 2 \ell_d^2 \log (\ell_d/a) F
(\Omega \gamma/\varepsilon_0)$. The full $B({\bf k})$ reflects the
ATGS instability: It is positive for large $\vert {\bf k} \vert$, but
negative below a critical wavenumber $k_c$, attaining a minimum at the
wavevectors
$\vert {\bf k} \vert \! \equiv \! k_\ast \! = \! 3/4 \, \alpha^2/\pi
\, (1 \! + \! \sigma)/(1 \! - \! \sigma) \, E/\gamma$,
where $B({\bf k})$ takes the value $- \gamma k_\ast^2 / 3$. In Eq.\
(\ref{lin_surf}) because of (even small) initial roughness and noise
in the  deposition and diffusion processes from this linear
instability random patterns of buckles with typical scale
$\lambda_{\ast} \! = \! 2 \pi / k_\ast$ emerge growing exponentially
in amplitude \cite{KR99} with a rate $1/\tau_\ast \! \equiv \! 2/3 \,
\ell_d^2 \log (\ell_d/a) F (\Omega \gamma/\varepsilon_0) k_\ast^4 \!
= \! \kappa k_\ast^4 / 3$.

So far a nonequilibirum dynamic equation for growing surfaces,
Eq.\ (\ref{final_lin}), has been derived perturbatively close to a
horizontal interface. Now it is compared to experimental results
\cite{MoskeHabil,Moske97,Moske99,Yang96,Lutt97}. The
instability has been carefully observed e.g.\ for Zr$_{\rm
65}$Al$_{\rm 7.5}$Cu$_{\rm 27.5}$ \cite{Moske97}, but the lateral
stress is documented only for related materials
\cite{MoskeHabil,Moske89}. The best test would be to measure both
simultaneously, but already with the given information an order of
magnitude estimate for $\lambda_{\ast}$ and $1/\tau_\ast$ is possible. 

Here are the experimental parameters: The elastic constants were given
above, the surface tension is $\gamma \! \approx \! 2$ J/m$^2$
\cite{MoskeHabil}, temperature $k_B T \! \approx \! 4 \! \times \!
10^{-21}$ J, growth velocity $F \! = \! 8$ \AA/s, atomic size $a \! = 
\! \Omega^{1/3} \! \! \approx \! 3$ \AA \cite{Moske97}. Order of
magnitude estimates are used for the time between adatom hop attempts,
$\tau_0 \! \approx \! 10^{-13}$ s \cite{SmilauerVvedensky}, and
surface energetics, $\varepsilon_0 \! \approx \! 10^{-19}$ J (about 1
eV). This yields a diffusion length $\ell_d \! \approx \! 1$ nm, which
suits well the observation that atoms move a few diameters before
incorporation \cite{Moske97,Geyer97}. Indeed this is smaller than the
average distance between mobile adatoms $\sqrt{\Omega/\rho_0} \!
\approx \! \ell_d \sqrt{D/F} \! \approx \! a \sqrt{\varepsilon_0/k_B
T}$ (D is calculated from the distribution of waiting times in
``unsticky'' sites \cite{Kehr}), so dimer formation is
suppressed. Given these values the theoretical predictions are
$\lambda_{\ast} \! \approx \! 25$ nm and $1/\tau_\ast \! \approx \!
10^{-2}$ 1/s.

\begin{figure}[htbp]
\narrowtext
\unitlength1cm
\begin{center}
   \begin{picture}(6,2.7)
      \includegraphics{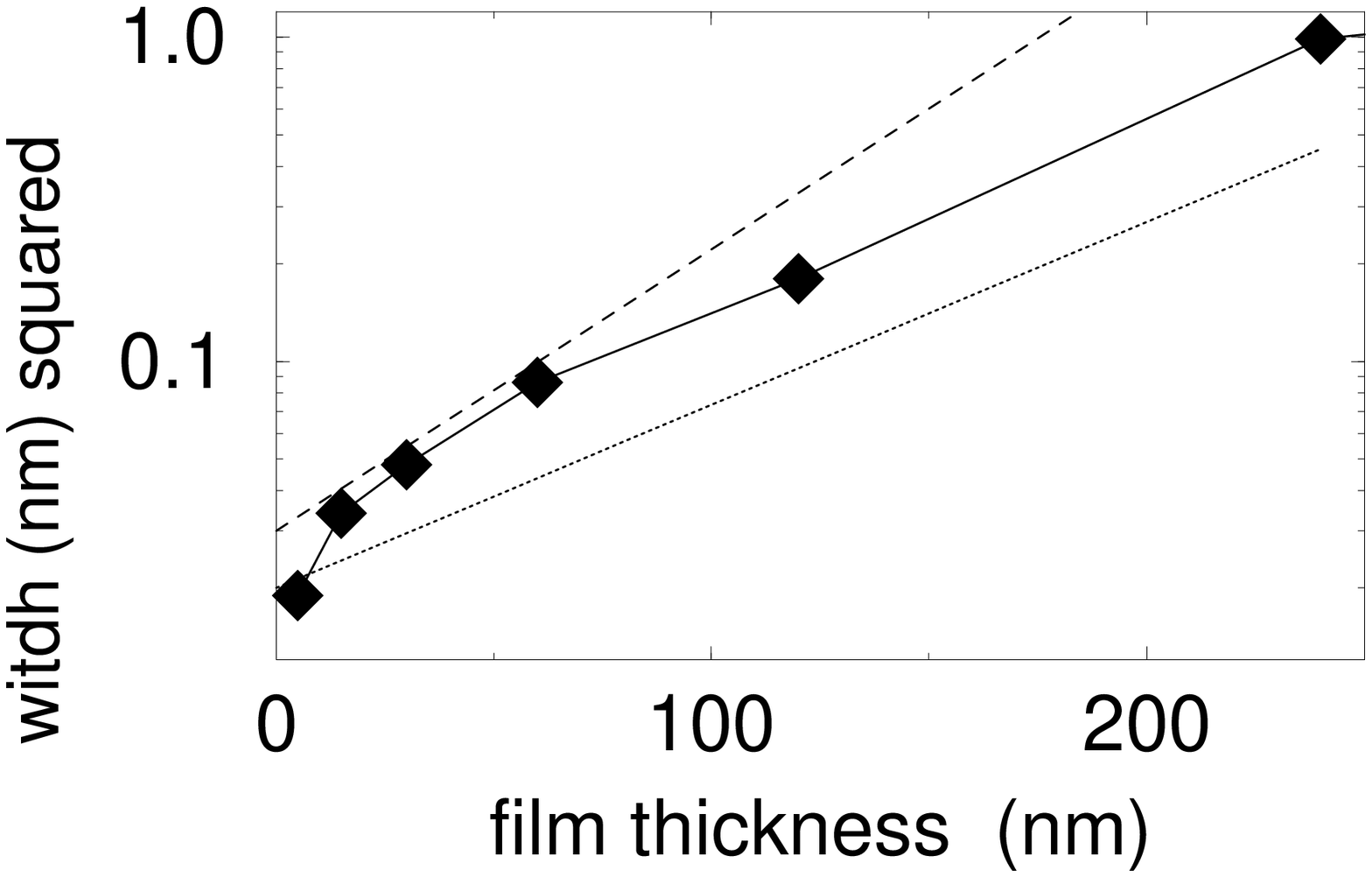}
      \includegraphics{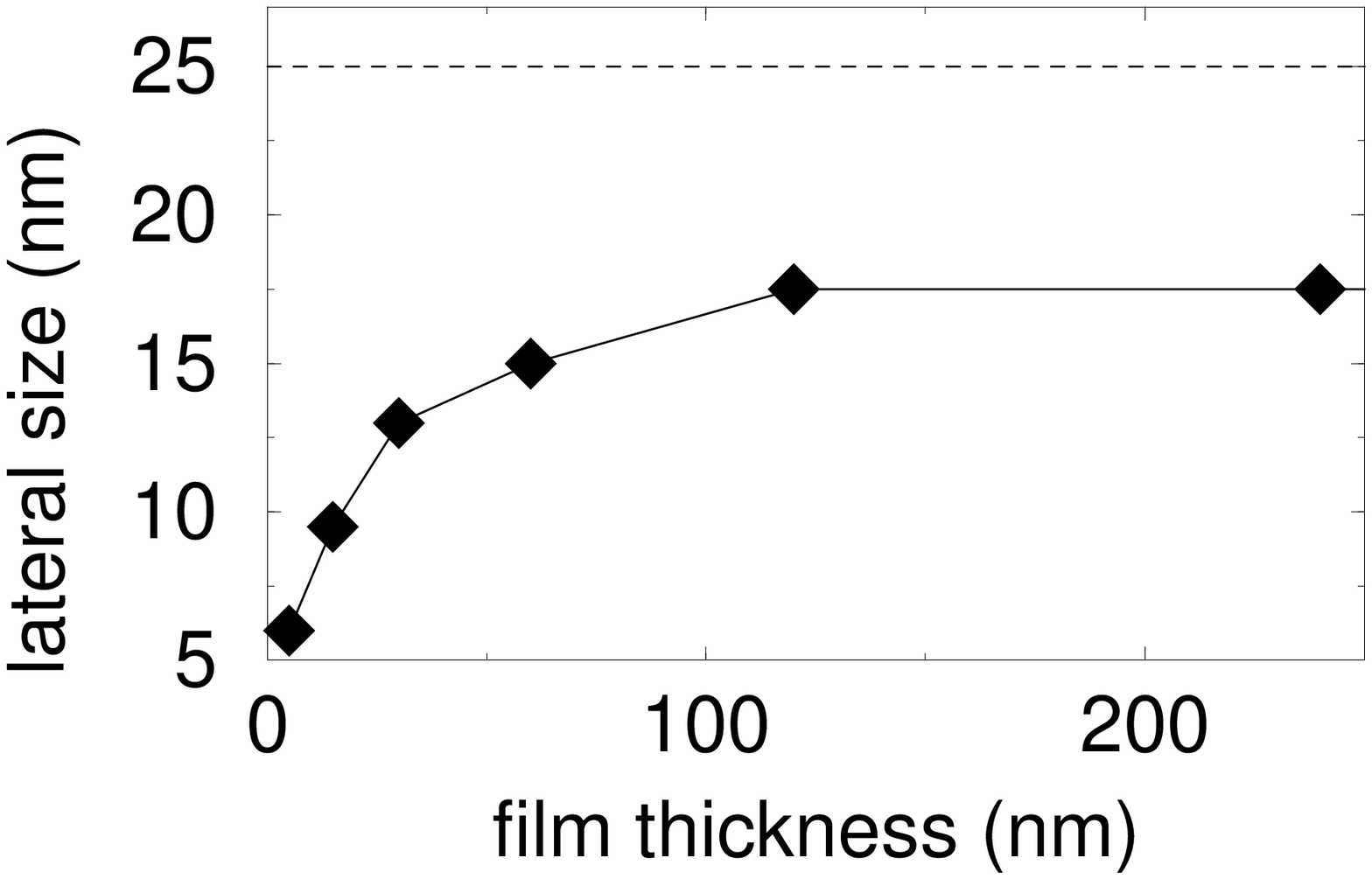}
   \end{picture}
\end{center}
\caption{Comparing $1/\tau_\ast$ and $\lambda_\ast$ (dashed lines) to
experiments \protect \cite{Moske97} (diamonds). Left: mean square
roughness, growing exponentially after thickness 30 nm. Dotted line
shows growth rate $6.5 \! \times \! 10^{-3}$ 1/s. Right: lateral buckle size.} 
\label{Vergleich}
\end{figure}

This fits fairly well to experimental observations, as illustrated in
Fig.\ \ref{Vergleich}. The growing film
develops buckles, which after a thickness of about 30 nm take a
constant lateral size of $R_c \! \approx \! 17$ nm. From 30 to 240 nm 
film thickness their vertical amplitude increases exponentially with
rate $6.5 \! \times \! 10^{-3}$ 1/s. Surprisingly this quantitative
expression of a linear growth instability is not adressed explicitly
in the original work \cite{Moske97}, where the authors focus on an
observed early time algebraic increase for both quantities. It is
caused by
kinetic roughening
of large $k$ modes at times before
$\tau_\ast$
\cite{Elasti}. Besides,
in experiments on related materials the lateral homogeneous strain
$\alpha$ reaches a constant level only after about 50 nm
film thickness \cite{MoskeHabil,Moske89}. So only then the linear
instability as described by Eq.\ (\ref{final_lin}) with $B({\bf k})$
constant in time becomes visible.

Some remarks comparing to different interpretations of the observed
instability \cite{Moske97,Moske99} are in order. First, the elastic
energy density in Eq.\ (\ref{E_el_misfit}) is calculated for film and
substrate of the same material \cite{LandauLifshitz,Elasti}. This does
apply to the experimental system, where the Zr alloy film was
deposited on a previous 100 nm thick layer of the same material. In
particular, effects of perfectly rigid substrates \cite{Spencer91}
won't be observed. Second, the partial relaxation of the film close to
the modulated surface will not produce a measurable relief of total
stress in the layer. Even close to the surface the stress is lowered
only by a factor $1 \! - \! O(\vert \nabla h \vert)$ \cite{Elasti} and
the method of substrate deformation measures only an average stress
across the whole film \cite{MoskeHabil}. Third, compression of convex
expansion of concave parts by surface tension should not be measurable
in the total stress changes, since positive and negative curvature
compensate each other. Besides it is only a minor effect (compare Eq.\
(\ref{E_el_surftens}) to (\ref{e_cap}) and
(\ref{E_el_misfit})). Forth, recently a local continuum equation with
linear terms $\partial_t h = (\nu k^2 - \kappa k^4) h$ has been fitted
to experimental results \cite{Moske99}. It would be interesting to see
whether a destabilizing term $\propto \! \vert {\bf k} \vert^3$ as in
Eq.\ (\ref{lin_surf}) can give a better description.

In conclusion, in this letter a theoretical framework for a continuum
thery of surface growth with diffusion has been constructed. As in the
fundamental lattice models \cite{WolfVillain,DasSarma} the basic
processes are particle deposition and diffusion until an energetically
favourable site is reached. Surface free energy stabilizes the
interface by a configuration dependent attachment rate. For amorphous
solids a mean field type of approach yields Eq.\ (\ref{inc_change}), a
configuration dependence through the energy density near the surface,
resulting in a nonequilibrium evolution equation
(\ref{final_lin}). Second, an experimentally observed growth
instability on amorphous films has been shown to be stress
induced. Its spatial properties can be explained by standard energy 
arguments of the Asaro-Tiller-Grinfeld-Srolovitz instability, its
temporal evolution needs the above nonequilibrium framework. Far from
equilibrium the instability is enhanced. A challenge is the extension
to crystal growth, where steps act as sticky sites and cannot be
treated in a simple mean field way, and where island nucleation
becomes important. Random nucleation leads to different step
configurations on maxima compared to minima, so it contributes to
$I_1({\bf k})$ which may be a way to understand the relation $\kappa
\! \propto \! \ell_D^4 F$ obtained from dimensional analysis
\cite{Politi99,PolitiVillain96}, where $\ell_D$ denotes the average
distance between island nuclei on crystal terraces, conceptually
different from $\ell_d$ here. Also the elastic interactions should be
worked out for crystals and for heteroepitaxy with different
rigidities of substrate and growing film, which would enable very
important applications \cite{Elasti}.

It is a pleasure to thank Joachim Krug for encouraging discussions and 
helpful comments. This work has been supported by the Academy of
Finland.

\end{multicols}

\begin{references}
\bibitem{Linear} Throughout this work dynamics are only considered to
linear order in $H$.
\bibitem{CommExp} Any other form of $n(\varepsilon/\varepsilon_0)$
gives results of the same order of magnitude, as long as mean and
width of the distribution are of order $\varepsilon_0$.
\end{references}
\end{document}